\documentclass[11pt]{article}
\usepackage{epsfig}
\if@twoside\oddsidemargin25mm
\else\oddsidemargin25mm\evensidemargin25mm\marginparwidth25mm\fi
\def\bpl{\Big(}
\def\bpr{\Big)}

\def\t{\theta}

\def\k{\kappa}

\def\bq{\begin{equation}}
\def\eq{\end{equation}}
\def\brr{\begin{eqnarray}}
\def\err{\end{eqnarray}}
\def\ba{\left(\begin{array}}
\def\ea{\end{array}\right)}

\newcommand{\dr}{\raise.3ex\hbox{$\stackrel{\leftarrow}{\partial }$}{}}
\newcommand{\dl}{\raise.3ex\hbox{$\stackrel{\rightarrow}{\partial}$}{}}

\newcommand{\ft}[2]{{\textstyle\frac{#1}{#2}}}

\begin{document}
%%%%%%
\renewcommand{\a}{\alpha}
\renewcommand{\b}{\beta}
\renewcommand{\c}{\gamma}
\renewcommand{\d}{\delta}
\newcommand{\pa}{\partial}
\newcommand{\g}{\gamma}
\newcommand{\G}{\Gamma}
\newcommand{\A}{\Alpha}
\newcommand{\B}{\Beta}
\newcommand{\D}{\Delta}
\newcommand{\e}{\epsilon}
\newcommand{\E}{\Epsilon}
\newcommand{\z}{\zeta}
\newcommand{\Z}{\Zeta}
\renewcommand{\l}{\lambda}
\renewcommand{\L}{\Lambda}
\newcommand{\La}{\Lambda}
\newcommand{\m}{\mu}
\newcommand{\M}{\Mu}
\newcommand{\n}{\nu}
\newcommand{\N}{\Nu}
\newcommand{\x}{\chi}
\newcommand{\X}{\Chi}
\newcommand{\p}{\pi}
\newcommand{\R}{\Rho}
\newcommand{\s}{\sigma}
\renewcommand{\S}{\Sigma}
\newcommand{\T}{\Tau}
\newcommand{\y}{\upsilon}
\newcommand{\Y}{\upsilon}
\renewcommand{\o}{\omega}
\newcommand{\q}{\theta}
\newcommand{\h}{\eta}
%%%%%
\begin{titlepage}
\begin{flushright} HUB-EP 98/23 \\[1mm] 
{\tt hep-th/9803254}
\end{flushright}
\vspace{16mm}
\begin{center}
{\LARGE\bf Confluences of Anomaly Freedom Requirements\\[.3in]
in {\it M}-theory}
\vfill
{\large Michael Faux
\footnote{faux@qft15.physik.hu-berlin.de}}\\
\vspace{7mm}
{\small
Humboldt-Universit\"at zu Berlin, Institut f\"ur Physik\\[1mm]
D-10115 Berlin, Germany\\[6pt]
 }
\end{center}
\vfill
\begin{center} {\bf Abstract} \\[3mm]
{\small
A topological fact about eleven dimensions is used to motivate a
potential new duality in {\it M}-theory.  We complete the discussion
of consistent limits of {\it M}-theory raised in a previous paper
\cite{faux1}, to include gravitational anomaly cancelation
and four-form flux quantization in the context of the 
$M^{10}\times S^1/Z_2$ compactifications. A suprise is found:
If one includes a $Z_2$ anomaly which exists only in $8k+3$ dimensions
then there are two distinct quantum limits, one related to each
of two equivalency classes of the orbifolds $M^{10}\times S^1/Z_2$.}  

\end{center}
\vfill
\flushleft{March 1998}
\end{titlepage}

\section{Introduction}
In $8k+3$ dimensions the sign of path integrals over
fermions coupled to gauge fields and gravity is potentially ambiguous.  
This has been used  in \cite{wittenflux} 
to prove a half-integer quantization on
period integrals of the four-form $G$ of {\it M}-theory
based on path integrals over worldvolume fermions
on the {\it M}-theory membrane
(which has a three-dimensional worldvolume so that $k=0$).  
The relevant ambiguity is resolved by the specific topology of spacetime.
Interestingly, the next higher dimension where this ambiguity could 
occur is eleven, where $k=1$. In this paper we present an intriguing
consequence of this potential ambiguity, making a case for 
yet a new duality in {\it M}-theory.

In the construction of the low-energy effective action for {\it M}-theory
compactified on $M^{10}\times S^1/Z_2$, there are {\it two} parameters
which remain unspecified by the minimal implementation of supersymmetry
(ie: without encorporating loop effects).
In this compactification, the primary 
goal is to couple eleven-dimensional supergravity to ten-dimensional vector 
multiplets propagating on each of the two hyperplanes fixed
by the $Z_2$ projection.  It is well known by now that the consistent 
coupling requires that each of these vector multiplets live in the 
adjoint representation of $E_8$.  This construction also represents
the strong-coupling limit of the $E_8\times E_8$ heterotic
string theory.  The two parameters which remain unfixed in the minimal
coupling are the dimensionless ratio $\nu\,\propto\,\l^6/\k^4$, where $\l$ is
the $E_8$ coupling constant and $\k$ is the eleven dimensional
gravitational coupling constant, and a particular angle which 
appears in the definition of the four-form field strength $G$.
The first parameter, and the puzzle regarding its resolution
was discussed in \cite{hw2}.  The second 
parameter was described in \cite{faux1, lu, conrad}, and
relates to the necessary coupling of $G$ to the Chern-Simons
forms associated with the $E_8$ gauge potentials.  
We call this second parameter $b$.  It reflects 
a freedom to partially manipulate the Chern-Simons coupling from components
$G_{(11)ABC}$ to components $G_{ABCD}$, where $A,B,C$ and $D$ do not
take the value $11$.

As explained in \cite{faux1} {\it both} of the parameters $\nu$ and $b$ 
are required to cancel all anomalies in the theory, and both parameters
are fixed by this requirement.  

There are three types of anomalies which we need to cancel.  The first
of these are the gauge anomalies, under which the $E_8\times E_8$
symmetery could be destroyed by quantum effects.  The second are
gravitational anomalies, under which eleven-dimensional
general coordinate invariance (or equivalently local Lorentz invariance)
could likewise be destroyed.  The third type of anomaly relates to
a topological quantization of $G$ necessary to ensure the consistency
of membranes in the quantum theory.  In some sense this 
third anomaly is related to the tensor symmetry under which the three-form
gauge potential transforms as $C\rightarrow C+d\Lambda$.

Each of the three anomalies described in the last paragraph imply
an independent constraint on the parameters $b$ and $\nu$.  As
described and proved in this paper, each of these constraints takes the
form of a cubic equation defining a locus of permitted values
of the pair $(b,\nu)$.  It is remarkable that these loci can intersect
at a common point.  It is even more remarkable that there are precisely
{\it two} such intersections.  There are two main purposes of this
paper.  The first is to derive the three cubic equations describing
the three independent anomaly cancelation requirements, and 
the second is to explain the relationship between the two intersection points.
The reader may benefit from taking a preliminary glance at the figure.

For the case of {\it M}-theory compactified on
$M^{10}\times S^1/Z_2$ the anomalies arise for the following 
reasons.  First of all, because of the $Z_2$ projection 
the spin 3/2 gravitino $\psi$ satisfies a constraint 
$\Gamma_{11}\psi\,|=\psi\,|$, valid precisely on the fixed points of the projection.
From the point of view of the ten-dimensional hyperplanes
defined by the $Z_2$ fixed points, this amounts to a chirality constraint.
As a consequence of this, the gravitino fields contribute to a
gravitational anomaly which is localized on the two fixed hyperplanes.
This can only be cancelled if vector multiplets propagate on
the fixed hyperplanes.  These supply spin 1/2 gauginos which,
in turn, further contribute to the gravitational
anomaly.  The net gravitational anomaly can be canceled only if the gauge 
group is $E_8$.  But at the same time, the $E_8$ transformations
are subject to a gauge  anomaly, also due to both the gravitino 
and the gauginos, which must be simultaneously canceled.

When we refer to the gravitational or gauge anomalies, 
we refer to the anomalous variation
of the quantum effective action $\d\Gamma$ attributable to 
one-loop. An important aspect of our analysis concerns 
the overall {\it sign} of this anomaly. 
Although the precise structure and the precise coefficients
which define $\d\Gamma$ are readily determined 
from established results, we indicate that the overall {\it sign} of the 
anomaly is not so readily obtained, particularly in eleven
dimensions, for reasons having to do with the ambiguity
described above in the introductory paragraph.  

Although the overall sign of the gravitational and gauge anomalies
{\it is} correlated with
the chirality of the gauginos (which are in-turn correlated 
with the chirality of the gravitino projection at the 
orbifold fixed points), this concern remains independent of the 
ambiguity which we are discussing, although in the end there
may be a connection.  

A conservative way to view the results of this paper is as an exploration
of the effects of the overall sign of the one-loop gauge
and gravitational anomalies on the supergravity construction
described in the second paragraph.  This analysis was motivated by an
observation that, curiously, there exists an elegant solution
to the anomaly cancelation requirements via a generalized 
Green-Schwarz mechanism for {\it either} of the the two naive choices
for the overall sign of this anomaly.  Since it is not a-priori
clear how this sign is fixed for the case of {\it M}-theory, 
since at the very least it involves a variety of deep issues 
from Euclideanization to regularization we leave this sign as 
an arbitrary parameter, and are encouraged by the observation
in \cite{wittenflux} that such an ambiguity is expected
for $8k+3$ dimensional path integrals.  We defer a more detailed
examination of the issue to a future publication.
However, the elegance of our results speaks for itself. 
With this caveat, we refer to a sign ambiguity on $\d\Gamma$
as the $Z_2$ anomaly. 

The essence of this $Z_2$ anomaly is that the path integral obtains a factor
$(-1)^\mu$, where $\mu$ is an integer determined by topological
considerations on the manifold.  For the case at hand, these would
include the $Z_2$ nature of our orbifold as well as the topology 
of $M^{10}$. Since $\mu$ necessarily falls into one of two
equivalency classes ``even" or ``odd", and since $\mu$ only
enters into our discussion as the factor $(-1)^\mu$, we can
without loss of generality restrict $\mu$ to take one of
two values, 0 or 1.  Furthermore, the path integration enters 
our discussion via its role in the computation of the anomaly.
The effect is that the expression $\d\Gamma$ which we would naively
use in determining the anomaly freedom of {\it M}-theory should be
replaced with $(-1)^\mu\d\Gamma$, where $\mu$ remains an unspecified
parameter which takes values 0 or 1 depending on yet-to-be-determined
topological properties of $M^{10}\times S^1/Z_2$.

The anomaly freedom requirements of {\it M}-theory were previously
analyzed for the choice $\mu=0$ by several authors 
\cite{lu, conrad, dealwis2, dm}.  In this paper we leave $\mu$
unspecified and find a more general set of constraints.  As described
above these distill into a system of three cubic equations
which permit exactly two simultaneous solutions, one corresponding to 
each of the two choices $\mu=0$ or $\mu=1$.  The unique point
corresponding to $\mu=0$ describes to a previous result
due to Lu in \cite{lu}, which requires $\l^6/\k^4=(4\pi)^5$.
The unique point corresponding to $\mu=1$ requires $\l^6/\k^4$ 
to be exactly 1/27 of the $\mu=0$ result.  It is nontrivial that 
either one of these cases should admit a solution, and quite
remarkable that there is exactly one solution for each case.
The possibility of a $Z_2$ anomaly highlights
a possible duality between compactifications with 
spacetime topology corresponding to one value $\mu$ with those 
corresponding to the opposite value of $\mu$.
The analysis of this paper involves a satisfying picture 
which includes both possibilities in a common framework.

Another remarkable aspect of {\it M}-theory is that its precepts offer
an a-priori explanation for all Green-Schwarz terms
necessary for gauge and gravitational anomaly cancelation.
By way of contrast, in the case of weakly-coupled string theory, 
the effective ten-dimensional supergravity theory 
has only the anomalies themselves to justify some of the 
Green-Schwarz terms.  In {\it M}-theory, the counter terms are
given by
\bq S_{GS}=-\frac{\sqrt{2}}{\k^2}\int_{M^{11}}\bpl\,
    C\wedge G\wedge G
    -\frac{1}{(4\pi)^3\,T_5}\,C\wedge X_8\,\bpr \,.
\label{gs}\eq
The first term is known from ancient history \cite{cjs}, as it is 
one of the minimal couplings required by supersymmetry in eleven
dimensions. However, the fundamental role of this term 
has only become evident in the last few years.  
The second term is required since {\it M}-theory contains fivebranes.  
The absence of worldvolume anomalies for the fivebrane requires this 
coupling. The eight-form $X_8$ is determined by this restriction
\cite{dlm, wittenfive}, and
is given by
\bq X_8=-\ft18{\rm tr}R^4+\ft{1}{32}({\rm tr}R^2)^2 \,,
\eq
where $R$ is the eleven-dimensional Riemann tensor expressed as a 
Lorentz-valued two-form.
The factor $T_5$ is the fivebrane tension which, for reasons explained
in section 3, is subject to a quantization rule given by
\bq T_5=\bpl\frac{\pi\,l}{2\k^4}\bpr^{1/3} \,,
\label{t5}\eq
where $l$ is an integer.  This quantization is related to the 
flux quantization of $G$.

In this paper we always take $\Gamma_{11}\psi\,|=\psi\,|$
(as opposed to $\Gamma_{11}\psi\,|=-\psi\,|$\,) so that
we do not include certain irrelevant sign ambiguities present in
\cite{faux1}.

In this paper we work exclusively in the upstairs approach.
Thus, we have a {\it boundary-free} eleven-dimensional orbifold
$M^{10}\times S^1/Z_2$, where the $Z_2$ projection acts 
on $x^{11}\in (-\pi,\pi)$ as $x^{11}\rightarrow -x^{11}$.
An alternative ``downstairs" point of view, in which the spacetime 
has boundaries has been discussed \cite{hw2, conrad, lu}.  
According to the proponents of that alternate approach, it is merely 
a matter of taste or convenience to employ one picture or the other.  
This being so, provided that all parameters are chosen
properly, it is clear that nothing is lost while clarity
is gained by simply sticking to one convention.   

This paper is structured as follows.

Section 2 comprises the main thrust of this paper.  Here we present a 
derivation of the two independent constraints imposed by gauge
anomaly freedom and gravitational anomaly freedom.
These as well as a third relation expressing the flux quantization
of $G$ (which is derived in section 3), comprise 
algebraic relations between the two parameters $\nu$ and
$b$ described above, as well as on a pair
of ostensibly independent topological numbers. One of these
is the number $l$ which defines the fivebrane tension, and the
other is a parameter $m$ related to the flux quantization of $G$.
The two conditions on $\nu$ and $b$ each consist of an equation cubic in 
the parameter $b$ and linear in the parameter $\nu$.  We discuss
the simultaneous solution of these constraints, showing that there are
precisely two solutions; the fivebrane tension is fixed, but
the quantized flux of $G$ can take one of two opposing values
which are correlated with the $Z_2$ anomaly described above.

Section 3 consists of a derivation of the flux quantization condition
for $G$ expressed as an algebraic relation between $b$ and $\nu$
in the form which was used in section 2.  This derivation is presented 
independently so as not to disturb the linear presentation in
section 2.  

Finally, section 4 summarizes what we have learned in this analysis,
and describes several directions of future research naturally
indicated by our results.

\section{{\it M}-theory on $M^{10}\times S^1/Z_2$}  
We are considering {\it M}-theory compactified on the orbifold 
$M^{10}\times S^1/Z_2$.  As is well known by now, this requires 
ten dimensional vector supermultiplets to propagate on 
each of the two ten-dimensional hyperplanes fixed by the orbifold 
projection.  Considerations based on gauging the superalgebra then
imply the following modification to the definition of the four-form
field strength,
\bq G=6dC-\ft16(2b+3\sqrt{2})\,\frac{\k^2}{\l^2}\,
    \sum_{i=1}^2\,\d(x^{11}-a_i)\,dx^{11}\wedge Q_{3\,(i)}^0
    -\ft16\,b\,\frac{\k^2}{\l^2}\,\sum_{i=1}^2
    \t(x^{11}-a_i)\,I_{4\,(i)}\,.
\label{gdef}\eq
In this expression the sums include contributions
from each of the two orbifold fixed 
points, which are defined by $x^{11}=a_i$ where $a_1=0$ and
$a_2=\pi$.  Note that the one-form gauge potentials $A_{(i)}$, and 
therefore the respective
two-form field strengths $F_{(i)}$,
are constrained to propagate on the fixed hyperplanes.
The Heavyside step function $\t(x^{11}-a_i)$ has the property that
$d\t(x^{11}-a_i)=2\d(x^{11}-a_i)\,dx^{11}$.
The parameter $b$ is a real parameter which was not described in the
original work \cite{hw2}.  As was described in \cite{faux1},
and also previously in \cite{dealwis2, conrad, lu} 
\footnote{The parameter $b$ in this paper is related to the parameter 
$\a$ in \cite{lu} via the following correspondence,\\
$b=-3(1+\a)/\sqrt{2}$.}
this parameter is necessary for full anomaly cancelation.  
Further, its value is
only selected by the requirement that quantum anomalies be absent
(which includes the flux quantization of $G$).
The four-form $I_{4\,(i)}$ is
defined by
\footnote{As is customary for $E_8$ valued
matrices, ${\rm tr}\equiv\ft{1}{30}Tr$, where ${\rm Tr}$ is the trace
in the adjoint representation.}
\brr I_{4\,(i)}&=&\ft12\,{\rm tr}R\wedge R
    -{\rm tr}F_{(i)}\wedge F_{(i)} \nonumber\\[.1in]
    &=& dQ_{3\,(i)}^0 \,,
\label{i4def}\err
where $Q_{3\,(i)}^0$ contains the Yang-Mills Chern-Simons three-form
associated with the gauge potential $A_{(i)}$ and also the
Lorentz Chern-Simons three-form associated with the spin connection.
Note that the presence of the $F_{(i)}\wedge F_{(i)}$ term
in $I_{4\,(i)}$ follows from minimally implementing supersymmetry.
Nicely, this contribution is independently
required for gauge anomaly cancelation.
These terms are required even in a classical theory in order to
resolve the gauge superalgebra between the bulk eleven-dimensional
supergravity fields and the $E_8$ fields propagating on the fixed
hyperplanes.  The $R\wedge R$ modifications are included to enable
cancelation of gravitational anomalies.  The factor of one
half that multiplies the $R\wedge R$ term arises, essentially, because
the gravitational anomaly is equally distributed between the two
fixed hyperplanes.

The modified definition for $G$ given in (\ref{gdef}) has as a consequence
the following Bianchi identity,
\bq dG=\frac{1}{\sqrt{2}}\,\frac{\k^2}{\l^2}\,\sum_{i=1}^2\,
    \d(x^{11}-a_i)\,dx^{11}\wedge I_{4\,(i)} \,.
\label{bianchi}\eq
Note that the parameter $b$ cancels when applying the exterior 
derivative to equation (\ref{gdef}).  Alternatively, one can view 
$b$ as parameterizing a {\it family} of solutions, given by
(\ref{gdef}) to
the bianchi identity (\ref{bianchi}).  When we include gauge anomaly
cancelation, the value of $b$ determines the ratio
$\l^6/\k^4$.  But one requires cancelation of gravitational anomalies
as well as a flux quantization rule on $G$ in order to pin down
the value of $b$ necessary for a consistent quantum theory. 

In what follows it is useful to define the dimensionless order parameter
\bq \nu=\frac{54\sqrt{2}}{(4\pi)^5}\,\frac{\l^6}{\k^4} \,.
\label{nudef}\eq
The numerical factor in this definition is chosen to optimally
simplify certain expression to follow, particularly 
(\ref{cond1})-(\ref{cond3}) below.

Gauge invariance of $G$ requires the three-form potential to transform
as
\bq \d C=-\ft{1}{36}\,(2b+3\sqrt{2})\,
    \frac{\k^2}{\l^2}\,\sum_{i=1}^2\,
    \d(x^{11}-a_i)\,dx^{11}\wedge Q_{2\,(i)}^1 \,,
\label{dc}\eq 
where $Q_{2\,(i)}^1$ is defined by the relation
$\d Q_{3\,(i)}=dQ_{2\,(i)}^1$.  The two-form $Q_{2\,(i)}^1$
is linear both in the transformation parameter for the $i$th $E_8$ factor and
in the transformation parameter for a local Lorentz transformation.
Thus, $C$ has special properties under each of these
transformations.

As a consequence of the above, the Green-Schwarz terms shown in
(\ref{gs}) transform as follows,
\brr \d S_{GS} &=& \frac{1}{12\,(4\pi)^5}\,
    \frac{1}{\nu}\,
    (2b+3\sqrt{2})\,b^2\,
    \sum_{i=1}^2\,
    \int_{M^{10}_i}Q_{2\,(i)}^1\wedge I_{4\,(i)}\wedge I_{4\,(i)}
    \nonumber\\
    & & 
    -\frac{1}{3(4\pi)^5}\,
    \frac{1}{(l\,\nu)^{1/3}}\,
    (2b+3\sqrt{2})\sum_{i=1}^2\,
    \int_{M^{10}_i}Q_{2\,(i)}^1\wedge X_8 \,.
\label{dgs}\err
In deriving this result we have applied (\ref{dc}) to (\ref{gdef}),
have used the property of the step function 
that $\t(x^{11}-a_i)^2=1$, and have used
the definitions (\ref{t5}) and (\ref{nudef}).  It is then 
straightforward to algebraically manipulate the coefficients
into the form shown in (\ref{dgs}).

The quantum effective action gives rise to another anomalous variation
due to the presence of certain one-loop diagrams.  These are
given by
\bq \d\Gamma=\frac{(-1)^\mu}{96\,(2\pi)^5}\,\sum_{i=1}^2\,\int_{M^{10}_i}
    Q_{2\,(i)}^1\wedge (\,\ft14 I_{4\,(i)}\wedge I_{4\,(i)}-X_8\,) \,.
\label{anom}\eq
This result is readily assembled from results in \cite{hw2, gsw, agw, agdpm}.
Commonly the anomaly is expressed in terms of a formal twelve-form,
which is a polynomial in traces over wedge products of the 
curvature two-form $R$ and the $E_8$ field strength two-form $F_{(i)}$.
This gives rise to the above expression when a relevant pair of
descent equations are employed.  The reason why the anomaly is
encoded in a twelve-form rather than a thirteen-form (thirteen being
$D+2$ for $D=11$) is that the anomaly has a fundamentally ten-dimensional
character.  This is because the anomaly is concentrated only
on the ten-dimensional fixed hyperplanes, since the fermions contributing
to the anomaly, consisting of the gauginos and notably the eleven-dimensional
gravitino, are only chiral in a ten-dimensional sense.  The descent
equations themselves give rise to ambiguities when they are
applied to the twelve-form, particularly if factorization
is implemented before the descent equations are solved.  These ambiguities
are resolved by enforcing a total permutation symmetry on the gauge
factors associated with the result.  This reflects the Bose symmetry
of the responsible loop diagrams.  Equation (\ref{anom}) represents
the consistent anomaly determined in this way.  As explained in the
introduction, the overall sign of the anomaly is the only undetermined factor,
due to the potential for a $Z_2$ anomaly in
eleven-dimensions.  This is relevant to the case at hand since the 
path integral describing the quantum {\it M}-theory {\it is} 
eleven-dimensional.

The variation of the full quantum effective action is given by the sum
of the quantum anomaly (\ref{anom}) and the variation of the
Green-Schwarz terms (\ref{gs}).
Amazingly, due to the factorization property evident in (\ref{anom}),
these two types of contributions have precisely the same form;
the fact that the quantum anomaly factorizes in the way shown in 
(\ref{anom}) is one of the miracles of {\it M}-theory.  
Thus, we may impose $\d S_{GS}+\d\Gamma=0$ to enforce anomaly
freedom.  This relation implies two independent restrictions,
one implying the cancelation of the $Q_2^1\wedge I_4\wedge I_4$
terms in each of (\ref{dgs}) and (\ref{anom}) and the other from
the similar cancelation of the $Q_2^1\wedge X_8$ terms. 

\pagebreak

After some simple and straightforward algebraic manipulations,
these two conditions can be expressed, respectively, as
\brr (2b+3\sqrt{2})\,b^2 &=& -(-1)^\mu\nu 
     \hspace{.4in} \longleftarrow {\rm Gauge\,Anomaly\,Cancelation}
     \label{cond1}\\[.1in]
     (2b+3\sqrt{2})^3 &=& -(-1)^\mu\,l\,\nu 
     \hspace{.33in} \longleftarrow {\rm Gravitational\,Anomaly\,Cancelation}
\label{cond2}\err
The first of these requirements coincides with the elimination of
gauge anomalies, and was described in \cite{faux1}.  
The second equation is the additional
requirement imposed by gravitational anomaly cancelation.
Notice that this equation represents a {\it family} of loci
parameterized by the integer $l$.  As we will demonstrate, 
flux quantization of $G$ uniquely selects the curve
described by $l=1$.  Flux quantization requires
\bq b^3=m\,\nu 
    \hspace{.3in} \longleftarrow {\rm Flux\,Quantization}
\label{cond3}\eq
where $m$ is another integer. The complete set of requirements necessary to 
fix $b$ and $\nu$ is given by the three cubic equations 
(\ref{cond1}), (\ref{cond2}) and (\ref{cond3}).

It is simple to see that only $l=1$ is permitted.  
This is done by replacing the first factor on the left-hand-side
of (\ref{cond1}) with $-(-1)^\mu\,(l\,\nu)^{1/3}$, which is implied by (\ref{cond2}). We then replace the remaining $b^2$ factor with 
$(m\,\nu)^{1/3}$, which is implied by (\ref{cond3}).
The factors of $-(-1)^\mu\,\nu$ 
cancel out, leaving us with the condition that
$lm^2=1$. Since $l$ and $m$ are integers, the only possibilities are
$(\,l\,,\,m\,)=(\,1\,,\,\pm 1\,)$.  
The three equations, (\ref{cond1}), (\ref{cond2}) and (\ref{cond3})
are presented as curves in the $(b,\nu)$ plane in
the figure.   The two solutions are then easily visualized.
Nevertheless, we will now derive the solutions algebraically.

With $l=1$, we can solve for
$b$ by equating the left hand sides of (\ref{cond1}) and (\ref{cond2}).  
The result of doing this is a cubic equation
which factorizes as
\bq (b+3\sqrt{2})(2b+3\sqrt{2})(b+\sqrt{2})=0 \,.
\eq
The roots are then apparent.  Thus, there are three possibilities,
$b=-3\sqrt{2}$, $b=-\ft32\sqrt{2}$ and $b=-\sqrt{2}$.  
The second of these possibilities, $b=-\ft32\sqrt{2}$
would imply that $\nu=0$, as is easily seen by plugging the result
into (\ref{cond1}) or (\ref{cond2}).  But this 
is disallowed, since $\nu$ is nonvanishing in the quantum
theory (reflecting the necessity for the fixed-point
gauge multiplets).  This leaves only two possibilities.  We examine these in turn.

\vspace{.1in}
\noindent
$\mu=0$:\\[1mm]
For the root $b=-3\sqrt{2}$, we can see from (\ref{cond2})
that $\nu= (-1)^\mu\,54\sqrt{2}$ and then from (\ref{cond3})
that $m=-(-1)^\mu$.  But $\nu$ is positive by its definition
(\ref{nudef}), so this solution is only valid for the case $\mu=0$.
This then implies that $m=-1$ and, using (\ref{nudef}) we determine
that 
\bq \frac{\l^6}{\k^4}=(4\pi)^5 \,.
\label{sol0}\eq
This is the result obtained in \cite{lu}, and is represented
by the lower of the two marked intersection points shown in the figure.
Note that unlike the counterpart solution for $\mu=1$, this solution
does not occur at any special point (ie: turning points or inflection
points) on any of the three curves representing the three anomaly 
freedom requirements.  
Without the possibility of the $Z_2$ anomaly in $8k+3$ dimensions
this would represent the sole possibility for a consistent
quantum theory.  We could then take (\ref{sol0}) as a defining
relation for the $M^{10}\times S^1/Z_2$ orbifolding of {\it M}-theory.
However, as we have discussed, there is another possibility.

\vspace{.1in}
\noindent
$\mu=1$:\\[1mm]
For the root $b=-\sqrt{2}$, we can see from (\ref{cond2})
that $\nu=-(-1)^\mu\,2\sqrt{2}$ and then from (\ref{cond3})
that $m=-(-1)^\mu$.  But, again, $\nu$ is positive by its
definition (\ref{nudef}), so this solution is only valid for the
case $\mu=1$.  This then implies that $m=+1$ and, using (\ref{nudef})
we determine that 
\bq \frac{\l^6}{\k^4}=\frac{1}{27}\,(4\pi)^5 \,.
\eq
This solution is new, and is represented by the upper of the two
marked intersection points shown in the figure.  Since we cannot
ignore the $Z_2$ anomaly in eleven dimensions this 
represents the lone alternative to the $\mu=0$ solution.  
On certain aesthetic grounds, this solution has an advantage 
over the alternative.  This is because this solution exists 
at the only special point for $\nu\ne 0$ on the curve representing
the gauge anomaly condition.

\begin{figure}
\begin{center}
\includegraphics[width=90mm,angle=-90]{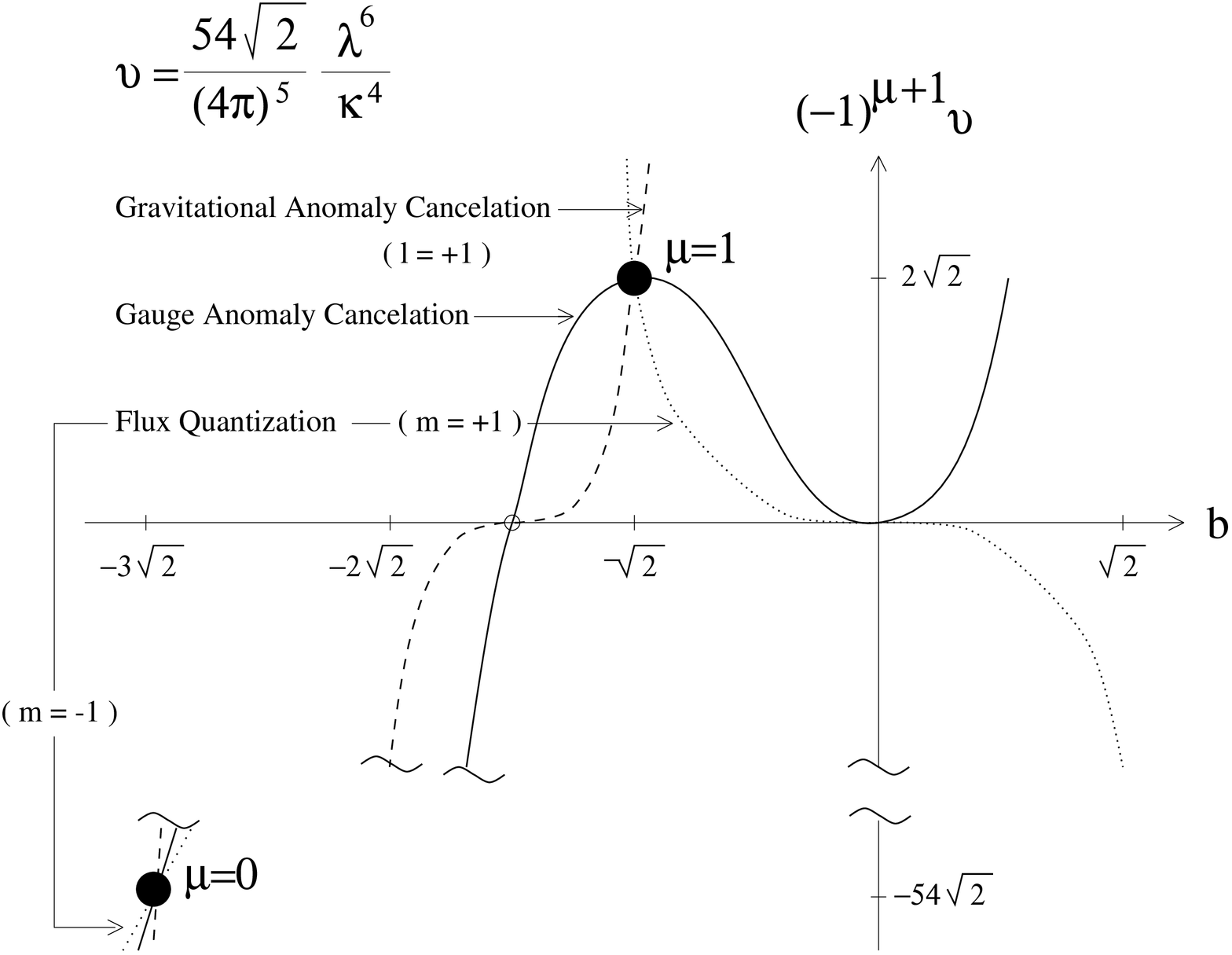}\\[.5cm]
\parbox{5in}{Figure: Loci of $(\,b\,,\,\nu\,)$
which permit cancelation of the indicated anomalies.
There are exactly two possible simultaneous solutions.
The upper one corresponds to the choice $\mu=1$, while
the lower one corresponds to $\mu=0$.  The $\mu=0$ choice was
discovered by Lu \cite{lu}.  The $\mu=1$ solution is new
to this paper.}
\end{center}
\end{figure}

\section{Flux Quantization}
The four-form $G$ obeys flux quantization such that, in 
appropriate units, its periods are half-integer
(ie: half of an odd integer).  This has been demonstrated by Witten
in \cite{wittenflux}, by using a little-known fact that there can be
a topologically-resolved sign ambiguity for path integrals 
over fermions in 8$k$+3 dimensions.  This applies to the {\it M}-theory
membrane (which has a three-dimensional worldvolume, or $k=0$), as
was demonstrated by Witten.
This discovery modified prior conventional wisdom 
that $G$ has integer periods
\footnote{Witten also offered a
heuristic justification based on the observation that
$\int G\sim 16\pi^2\int(F\wedge F-\ft12 R\wedge R)$
can be {\it either} integer or half-integer
depending on global properties of spacetime.  
So, if one could find a 
region of space over which a cycle $\int G$ were half-integer,
one could then extrapolate this local knowledge to infer 
something about the topology of remote regions of spacetime.  
He argued that this would be counterintuitive.}. 
To derive the precise rule requires a membrane worldvolume action as a 
starting point.  With $T_2$ defined as a relevant membrane tension
parameter, this quantization reads
\bq T_2\,\int G=\pi\,n
\label{fluxq}\eq
where $n$ is an odd integer, and the integral is over any four-cycle
in the universe.  To resolve this constraint in
terms of factors and parameters of the eleven-dimensional
supergravity requires comparison with the supergravity field equations,
obtained by the variation of
the $D=11$ supergravity action given in \cite{faux1}.
Enforcing consistency between these and the above flux quantization rule
implies a related quantization of the membrane tension, given by
\bq T_2=\frac{1}{\sqrt{2}}\bpl\,\frac{2\,\pi^2}{\k^2\,l}\,\bpr^{1/3}
\label{ten2}\eq
where $l$ is an integer.

The {\it M}-theory membrane is dual to the {\it M}-theory fivebrane.  As a 
consequence, the respective brane tensions are subject to a Dirac 
quantization rule.  For the conventions which we have chosen,
particularly the choice of factors in the supergravity action,
and the choice of membrane action, reflected in the 
quantization rule (\ref{fluxq}), this Dirac quantization 
requires that $\sqrt{2}\k^2T_2T_5/\pi$ is an integer.  
We can then use this relation to rewrite (\ref{ten2}) as 
\bq T_5=\bpl\frac{\pi\,l}{2\k^4}\,\bpr^{1/3} \,,
\label{ten5}\eq
which is the same as equation (\ref{t5}).
The issue of brane quantization is discussed in more detail
in \cite{dlm, lu}, and a relationship of these relations to
$D$-branes is discussed in \cite{dealwisb}.
Necessarily the identical quantization 
is obtained from analysis of the consistent fivebrane 
worldvolume theory.

As discussed in \cite{dlm}, stability of membranes and fivebranes 
requires that the respective brane tensions (and therefore 
the integer $l$) be necessarily positive.  It was pointed out
by Lu in \cite{lu} that this is automatically ensured for the
case $\mu=0$.  In section 2 we showed that $l=+1$ for any
consistent theory regardless of the value of $\mu$.  For this reason
we do not concern ourselves further with this restriction.

We can rewrite (\ref{fluxq}) in a more useful form by substituting
(\ref{ten2}).  The integer $l$ and the odd integer $n$ then combine
as $m\equiv l\,n^3$, which is an integer.
The result of this simple reorganizing is 
\bq \int G =\bpl\,\sqrt{2}\pi\,\k^2m\,\bpr^{1/3} \,.
\label{fluxnew}\eq
It is in this form that the half-integral flux quantization rule is
most useful to us.  It is apparent, from the definition (\ref{gdef})
that this provides another constraint on the parameters
$b$ and $\nu$.  We presently examine this in detail.

\vspace{.1in}
\noindent
{\it Implications of Flux Quantization}:\\[1mm]
We wish to derive the implications of the flux quantization
(\ref{fluxnew}) on the parameters described above, particularly 
the parameter $b$ and the two coupling constants $\k$ and $\l$.
These latter two combine to the dimensionless parameter $\nu$
defined in (\ref{nudef}).  

We consider four-cycles which have no extension in the 
eleventh dimension.
Subtleties regarding the irregular functions are avoided by
considering cycles removed from the fixed hyperplanes.
Over such cycles, only the final term 
in (\ref{gdef}) contributes to integrals of $G$.  The step
function in this term factors out of the integration and contributes
a sign which we safely suppress in this section (and {\it only} in 
this section). This is permitted because we are only interested in the 
integrality of factors, so overall signs are irrelevant.  
At these points, the $R\wedge R$ terms of $I_{4\,(i)}$, defined
in (\ref{i4def}), are nonvanishing, so that the integrand
does not vanish identically.  In fact, {\it two} factors of $\ft12\,{\rm tr}\,R\wedge R$
contribute to the integral in question: one from each of the
two terms in the sum  
\footnote{This is because in the definition of
$I_{4\,(i)}$ (ie: equation (\ref{i4def}))
only the $F_{(i)}\wedge F_{(i)}$ term is actually localized on the 
$i$th fixed hyperplane; the $R\wedge R$ term remains nonvanishing throughout 
the bulk.}.  Finally, $\int R\wedge R=16\pi^2 a$, where $a$ is an integer.
Assembling this information, we determine that, based on the definition
of $G$ (ie: equation (\ref{gdef})), for the 
periods which we are considering, 
\bq \int G=\ft83\,b\,\frac{\k^2}{\l^2}\,\pi^2\,a\,.
\label{period}\eq
To obtain the new restriction on $b$, we equate the right hand sides of 
(\ref{period}) and (\ref{fluxnew}).  After a small amount of algebra,
and using again the definition of the parameter $\nu$
(ie: equation (\ref{nudef})), we find
$a\,b=(\nu\,m\,)^{1/3}$.  The strongest restriction
comes with the case $a=1$ since any higher integer choice for 
$a$ can be compensated by a scaling of $m$.  In other words, 
for {\it any} value of $a$ there must exist {\it some} integer $m$ 
for which this restriction is valid.  But, if there exists an appropriate $m$
for the choice $a=1$ then there necessarily exists an analogous
$m$ for any other choice of $a$. 
Thus, without loss of generality we take $a=1$, so that
\bq b^3=\nu\,m
\label{bflux}\eq 
where $m$ is some integer.  We reiterate that signs have been ignored
in this discussion. 

In principle, other restrictions would follow from applying the flux 
quantization to other four-cycles, for instance, cycles which live
entirely on the fixed hyperplanes, or cycles which have some extent
in the eleventh dimension.  These do not pose additional
restrictions, however.

This constitutes a proof of (\ref{cond3}), which is the same as
(\ref{bflux}).  This is Witten's half-integral flux quantization
expressed in terms of the parameters $b$ and $\nu$ discussed
above.  As for other anomaly freedom requirements, the requirement
of flux quantization also translates to a restriction of $(b, \nu)$
to a locus defined by a cubic equation, (\ref{bflux}).  It is
nontrivial, as well as intriguing, that this locus, and the two
ostensibly independent loci corresponding to gauge anomaly
cancelation and gravitational anomaly cancelation can be arranged
to intersect at a common point.  The fact that they actually have 
exactly two intersections begs the question of the relation between
the two points.  This issue seems resolved by the eleven-dimensional
$Z_2$ anomaly as discussed above. 
 
\section{Conclusions}
The potential for a $Z_2$ anomaly in eleven dimensions gives rise to 
a pair of distinct classes of consistent low-energy constructions of 
$M^{10}\times S^1/Z_2$ orbifoldings of {\it M}-theory, with 
distinct restrictions on the respective coupling constants.
In each case the gauge coupling constant is proportional to the 
gravitational coupling constant to the 2/3 power, but with different coefficients. Only one of these two cases has been previously discovered.  
Topology necessarily distinguishes which one of these is appropriate 
for a given compactification.  

The $Z_2$ anomaly was invoked in \cite{wittenflux} in order to
prove the half-integral flux quantization of the four-form $G$.
In that case it was the {\it three}-dimensional worldvolume
of the {\it M}-theory membrane which exhibited the anomaly
in question, and the analog of the parameter $\mu$ described
in this paper was related to a particular characteristic class
associated with spin bundles on the world-volume.  
It is necessary to analogously resolve the precise topological definition 
of $\mu$ in order to understand further the implications of the
results described in this paper.  That analysis represents the 
obvious extension to this paper.

There are three possibilities.  

In the first case, the topology of $M^{10}\times S^1/Z_2$
would permit either $\mu=0$ or $\mu=1$ depending on the characteristics
of a given $M^{10}$.  In such a scenario the relevant parameters
$\nu$ defined in (\ref{nudef}) which defines the gauge coupling
in terms of the gravitational coupling, and perhaps more
relevantly $b$ which defines the Chern-Simons couplings to 
the four-form $G$, will differ depending on the value of $\mu$.
It would then become important to explore what differences 
this would imply for the compactified physics
\cite{ns, low, dg, noy, aq}, and if these
differences have some analog in the case of the weakly coupled
heterotic string, perhaps even implications for mirror symmetry.
This possibility is attractive because the same exotic
phenomenon, that of a $Z_2$ anomaly which can only exist in
$8k+3$ dimensions, would have implications to {\it M}-theory
coming from both the $k=0$ case and from the $k=1$ case, the former related
to the membrane worldvolume and the latter from the eleven-dimensional
spacetime.  

In the second case, the topology of $M^{10}\times S^1/Z_2$ would 
necessarily imply that $\mu=0$.  Equivalently, there may be an
argument to suggest that the ambiguity at the heart of this
analysis is not relevant to {\it M}-theory.  This would clearly be
the least interesting of the three possibilities.  It would imply that 
the previous result of Lu \cite{lu} represents the final word
on the issue of parameter identification in the context of
anomaly cancelation for {\it M}-theory on $M^{10}\times S^2/Z_2$
orbifolds.

In the third case, the topology of $M^{10}\times S^1/Z_2$
would necessarily imply that $\mu=1$.  This would then 
require a revision in our understanding of parameter
identification  in the context of anomaly cancelation for
{\it M}-theory on $M^{10}\times S^1/Z_2$, and would represent
an interesting application of a somewhat exotic agent, the 
$8k+3$ dimensional $Z_2$ anomaly.


\begin{thebibliography}{99}
%
\bibitem{faux1}
M.Faux,
{\it New Consistent Limits of M-theory},
hep-th/9801204.
%
\bibitem{wittenflux}
E.Witten
{\it On Flux Quantization in {\it M}-Theory and the Effective Action},\\
J. Geom. Phys. 22 (1977) 1-13,
hep-th/9609122.
%
\bibitem{hw2}
P.Ho{\v r}ava and E.Witten, 
{\it Eleven-Dimensional Supergravity on a Manifold
with Boundary},\\
Nucl. Phys. {\bf B}475 (1996) 94-114,
hep-th/9603142.
%
\bibitem{wittenfive}
E.Witten,
{\it Five-Branes and {\it M}-Theory on an Orbifold},\\
Nucl. Phys. {\bf B}463 (1996) 383-397,
hep-th/9512219.
%
\bibitem{dlm}
M.J.Duff, J.T.Liu and R.Minasian,
{\it Eleven Dimensional Origin of String/String Duality:\\ 
A One Loop Test},
Nucl. Phys. {\bf B}452 (1995) 261-282. 
hep-th/0906126.
%
\bibitem{lu}
J.X.Lu,
{\it Remarks on M-Theory Coupling Constants
and M-Brane Tension Quantizations},
hep-th/9711014.
%
\bibitem{conrad}
J.O.Conrad,
{\it Brane tensions and coupling constants
from within M-theory},
hep-th/9708031.
%
\bibitem{dealwis2}
S.de Alwis,
{\it Anomaly Cancelation in M-theory},
Phys. Lett. 392B (1996) 332.
%
\bibitem{dealwisb}
S.P.de Alwis,
{\it Coupling of branes and normalization of effective actions\\
in string/M-theory},
Phys. Rev. D56 (1997) 7963-7977,
hep-th/9705139.
%
\bibitem{dm}
E.Dudas and J.Mourad,
{\it On the strongly coupled heterotic string},\\
Phys. Lett. B400 (1997) 71-79,
hep-th 9701048.
%
\bibitem{cjs}
E.Cremmer, B.Julia and J.Scherk,
{\it Supergravity Theory in Eleven Dimensions},\\
Phys. Lett. 76B (1978) 409.
%
\bibitem{gsw}
M.Green, J.Schwarz, and E.Witten,
{\it Superstring Theory},\\
Cambridge University Press, 1987.
%
\bibitem{agw}
L.Alvarex-Gaum{\'e} and E.Witten,
{\it Gravitational Anomalies},\\
Nucl. Phys. {\bf B}234 (1983) 269-330.
%
\bibitem{agdpm}
L.Alvarez-Gaum{\'e}, S.Della Pietra, and G.Moore,
{\it Anomalies and Odd Dimensions},\\
Ann. Phys. 163 (1985) 288-317.
%
\bibitem{ns}
H.P.Nilles and S.Stieberger, 
{\it String-Unification, Universal One-Loop Corrections\\
and Strongly Coupled Heterotic String Theory},\\
Nucl. Phys. {\bf B}499 (1997) 3-28,
hep-th/9702110.
%
\bibitem{low}
A.Lukas, B.A.Ovrut and D.Waldram, 
{\it On the Four-Dimensional Effective Action\\
of Strongly Coupled Heterotic String Theory},
hep-th/9710208.
%
\bibitem{dg}
E.Dudas and C.Grojean,
{\it Four-Dimensional M-theory and supersymmetry breaking},\\
Nucl. Phys. {\bf B}507 (1997) 553-570,
hep-th/9704177.
%
\bibitem{noy}
H.P.Nilles, M.Olechowski and M.Yamaguchi,
{\it Supersymmetry Breaking\\
and Soft Terms in M-Theory},
Phys. Lett. B415 (1997) 24-30,\\
hep-th/9707143.
%
\bibitem{aq}
I.Antoniadis and M.Quir{\'o}s,
{\it Supersymmetry breaking in M-theory},\\
Nucl. Phys. Proc. Suppl. 62 (1998) 312-320,
hep-th/9709023.
%
\end{thebibliography}
\end{document}